\documentclass[aps, prd, floats, floatfix, twocolumn, superscriptaddress, nofootinbib, showpacs, letterpaper, 10pt, preprintnumbers, notitlepage]{revtex4-1}

\usepackage{graphicx}
\usepackage{amsmath}
\usepackage{amsfonts}
\usepackage{amssymb}
\usepackage{amstext}
\usepackage{tensor}
\usepackage{fullpage}
\usepackage[english]{babel}
\usepackage{blindtext}
\usepackage{helvet}
\usepackage{cancel}
\usepackage{microtype}
\usepackage[pdftex,
  pdftitle={Mass of a Patch of an FRW Universe},
  pdfauthor={Sarah R. Geller, Jolyon K. Bloomfield, Alan H. Guth},
  bookmarks,bookmarksopen=false,
  pdfstartview={FitH},
  linktocpage=true
]{hyperref}

\usepackage{mathrsfs}
\usepackage{mathtools}
\usepackage{euscript} 

\begin{document}

\title{Mass of a Patch of an FRW Universe}

\author{Sarah R. Geller}
\email{sgeller@mit.edu}
\author{Jolyon K. Bloomfield}
\email{jolyon@mit.edu}
\author{Alan H. Guth}
\email{guth@ctp.mit.edu}
\affiliation{Center for Theoretical Physics, Laboratory for Nuclear Science, and Department of Physics, Massachusetts Institute of Technology, Cambridge, MA 02139, USA}

\date{\today}

\preprint{MIT-CTP/4971}

\begin{abstract}
	In 1963, Zel'dovich devised a method to define the mass of a closed Friedmann-Robertson-Walker (FRW) universe, showing that by this definition it is exactly zero. Rounding out this result, we show that the masses of flat and open universes are (unsurprisingly) divergent. We also present closed-form solutions for the trajectory of the boundary of a finite spherical patch of homogeneous pressureless dust for each class of curvature, exploring the dynamics of the boundary in detail. In all cases, the FRW patch emerges from a white hole. In the closed case, the patch expands to a maximum radius before contracting and entering a black hole, while flat and open FRW patches expand without bound. We compare our results to the classical expectations of Newtonian cosmology, showing that for small radii the Newtonian energy gives the leading correction to the rest mass energy. 
\end{abstract}


\maketitle

\section{Introduction}\label{sec:introduction}

Relativists have long debated the possibility of defining a meaningful expression for the total relativistic energy of an arbitrary curved spacetime. Existing formalisms for calculating total relativistic energy, such as the Arnowitt-Deser-Misner (ADM) energy, are applicable only to spacetime geometries which are asymptotically flat. In 1963, Yakov Zel'dovich devised a method for computing the total relativistic mass of a closed universe described by the Friedmann-Robertson-Walker (FRW) metric \cite{zeldovichppr, zeldovich}. Zel'dovich considered a finite spherically symmetric spatial region of an FRW universe filled with dust, surrounded with a vacuum described by the Schwarzschild metric. The result is a spatial region of FRW which transitions smoothly to an asymptotically flat exterior region such that spherical symmetry is preserved throughout. By considering the Schwarzschild mass of the enclosed region as its boundary is extended to include the entire closed universe, Zeldovich showed that the total mass of the enclosed patch vanishes in this limit. Three years after Zel'dovich's initial calculation, W. Israel published the well-known junction conditions, characterizing the conditions under which two geometrically distinct spacetime regions can be joined along a mutual boundary \cite{Israel:1967zz}. 

In this paper, we extend the results of Zel'dovich by calculating the total mass of a patch of a matter-dominated FRW spacetime of arbitrary curvature. We find that the total mass of an open patch of FRW spacetime diverges exponentially as the radius of the boundary of the patch is taken to infinity, whilst the mass of a flat FRW patch diverges as the radius cubed. A closed FRW patch is found to have vanishing mass in the limit as its boundary is taken the include the whole universe, confirming the results of Zel'dovich. We then describe the dynamics of the boundary of such an FRW patch, computing analytic expressions for the trajectory of the boundary in Schwarzschild coordinates. This is very similar to Oppenheimer-Snyder collapse \cite{oppenheimersnyder} in reverse.

In Section II of this paper, we construct the model for an FRW patch containing a pressureless dust joined to an asymptotically flat and spherically symmetric external region, and find a general expression for the total mass of such a patch as a function of its boundary radius. We show that the masses of open and flat FRW universes diverge as the size of the patch increases. In Section III, we explore the physical interpretation of the total relativistic mass, and compare our result at small circumferential radius to the classical prediction of Newtonian cosmology. Finally, in Section IV, we plot the trajectory of the boundary of the patch in Kruskal and Penrose diagrams.

\section{Constructing the Model}\label{sec:constructing_the_model}

We consider a matter-dominated (i.e., zero pressure) FRW universe, which has the property that each particle travels on a geodesic. We imagine identifying a spherical patch of this spacetime, which expands with the universe (i.e. is ``comoving'') so that the particles on the edge of the region stay on the edge of the region, with no particles crossing the boundary. We then imagine removing this region from the full spacetime, and inserting it into a space that is completely empty outside the FRW patch. The region outside the FRW patch is then described by a Schwarzschild metric. The mass of the FRW patch can then be identified as the mass appearing in the Schwarzschild metric, corresponding to the ADM mass of the composite space. 

The two regions need to be ``glued'' together along their common boundary, and the Israel junction conditions specify the conditions under which such gluing is consistent with Einstein's equations. These conditions require the induced metric on the boundary to be the same on both sides, and the extrinsic curvature tensors to also agree (as we have no surface stress-energy). This gluing can be valid for all time only if the FRW content is a pressureless dust, as we are assuming. Otherwise, a particle on the boundary will experience a (singular) pressure gradient, since there is no matter just beyond the boundary, so its trajectory will be radically altered. 

To fix some notation, let ${\cal M}^-$ refer to the FRW region, ${\cal M}^+$ refer to the Schwarzschild region, and $\Sigma^\pm$ refer to the two sides of the boundary between the two. Throughout this calculation, we use units in which $G = c = 1$. The gluing formalism presented here closely follows that presented by Poisson \cite{Poisson}.

\subsection{FRW Patch}\label{ssub:frw_universe}

The FRW metric describes a homogeneous and isotropic universe whose expansion is governed by a scale factor $a$. In (hyper)spherical coordinates $\xi^{\mu}=(\eta, \chi, \theta, \phi)$, the spacetime interval is
\begin{align}\label{eq:FRW}
ds^2=g_{\mu\nu}d\xi^{\mu}d\xi^{\nu}=a^2(\eta)[-d\eta^2+d\chi^2+S_k^2(\chi) d\Omega^2]
\end{align}
where $d\Omega^2\equiv d\theta^2+\sin^2\theta d\phi^2$ is the metric on the two-sphere and
\begin{align}
	\label{frwskofchi}
S_k(\chi)=\begin{cases}
\sin(\chi), &\text{ if } k=+1 \,\,(\text{closed})\\
\chi, &\text{ if } k=0 \,\,(\text{flat})\\
\sinh(\chi), &\text{ if }\,\, k=-1 \,\,(\text{open})
\end{cases}
\end{align}
where the following relation holds for all k:
\begin{align}\label{Srelations}
S'^2_{k}(\chi)+k S_{k}^2(\chi)=1.
\end{align}
Note that we use conformal time, as it simplifies later calculations, and that the scale factor $a(\eta)$ has dimensions of length.

Using the metric \eqref{eq:FRW} in the Einstein field equations
\begin{align}
G_{\mu \nu} = 8 \pi T_{\mu \nu},
\end{align}
with a perfect fluid stress-energy tensor
\begin{align}
T_{\mu \nu} = (\rho + P) u_\mu u_\nu + P g_{\mu \nu},
\end{align}
with energy density $\rho$, pressure $P$, and velocity vector $u^\mu = (1/a, \vec{0})$, yields the Friedmann equation
\begin{align}\label{friedmann}
\left(\frac{\dot{a}}{a}\right)^2 = \frac{8 \pi \rho a^2}{3} - k,
\end{align}
where we use overdots to indicate derivatives with respect to $\eta$. Conservation of stress-energy $\nabla_\mu T^{\mu \nu} = 0$ yields the continuity equation
\begin{align}
\dot{\rho} = - 3 \frac{\dot{a}}{a} (\rho + P).
\end{align}
Solutions for equations of state of the form $P=w\rho$, where w is constant, are presented in Appendix \ref{app:friedmann}.

To describe the FRW patch, we restrict the FRW manifold to $\chi \le \chi_0$, placing the boundary $\Sigma^-$ at $\chi = \chi_0$ with constant $\chi_0$. By construction, $\Sigma^-$ is an embedded hypersurface that retains the spherical symmetry of the FRW spacetime. The natural coordinates on $\Sigma^-$ are $y^i = (\eta, \theta, \phi)$, in direct correspondence with the three bulk coordinates. The projection tensor
\begin{align}
\tensor{e}{^\mu_i} = \frac{\partial \xi^\mu}{\partial y^i}
\end{align}
is then straightforward to compute. The induced metric on the boundary is given by
\begin{align}\label{induced}
d\sigma_-^2 = h_{ij} dy^i dy^j = a^2(\eta) \left[- d\eta^2 + S_k^2(\chi_0) d\Omega^2 \right].
\end{align}

We construct a unit normal to the boundary as $\vec{n} = \partial_\chi / a(\eta)$, pointing outwards. The corresponding one-form is $\tilde{n} = a(\eta) d\chi$. As required for a normal, $\tensor{e}{^\mu_i} n_\mu = 0$. The extrinsic curvature tensor on the boundary is given by
\begin{align}
K^-_{ij} = \tensor{e}{^\mu_i} \tensor{e}{^\nu_j} \nabla_\mu n_\nu.
\end{align}
The components can be rapidly calculated as
\begin{subequations}\label{frwextrinsic}
\begin{align}
K^-_{\eta \eta} &= K^-_{\eta \theta} = K^-_{\eta \phi} = K^-_{\theta \phi} = 0
\\
K^-_{\theta \theta} &= a(\eta) S_k(\chi_0) S'_k(\chi_0)
\\
K^-_{\phi \phi} &= \sin^2(\theta) K^-_{\theta \theta}.
\end{align}
\end{subequations}
These results are independently derived using a different formalism in Appendix \ref{app:extrinsic}.

\subsection{Schwarzschild Patch}\label{sub:schwarzschild}

The Schwarzschild metric provides the unique spherically-symmetric vacuum solution to the Einstein field equations. In Schwarzschild coordinates, with the coordinate chart $x^{\mu}=(t,r,\theta,\phi)$, the metric is written
\begin{align}\label{schw}
ds^2 = -f(r)dt^2+\frac{1}{f(r)}dr^2+r^2d\Omega^2
\end{align}
where $f(r) = 1 - 2M/r$. The constant M is identified as the mass of the gravitating body, in our case, the FRW patch.

We plan to glue the FRW patch to an outer region of Schwarzschild spacetime, a region consisting of Schwarzschild spacetime for all values of $r>R(\eta)$, where $R(\eta)$ is yet to be determined. The coordinates of the boundary have already been specified as $y^i = (\eta, \theta, \phi)$, but we need to identify the Schwarzschild coordinates for such points. We describe the boundary $\Sigma^+$ as the set of points $x^\mu = (T(\eta), R(\eta), \theta, \phi)$, where $T(\eta)$ and $R(\eta)$ are functions that need to be determined.

The induced metric on the boundary is given by
\begin{align}\label{inducedplus}
d\sigma_+^2 = \left(-f(R) \dot{T}^2 + \frac{1}{f(R)} \dot{R}^2\right) d\eta^2 + R^2 d\Omega^2
\end{align}
where overdots once again indicate derivatives with respect to $\eta$, the conformal time on the boundary. We now employ the first Israel junction condition, which specifies that the induced metric on both sides of the boundary must be identical: $d\sigma_+^2 = d\sigma_-^2$. Comparing Eqs.~\eqref{induced} and \eqref{inducedplus}, we obtain
\begin{align}
R(\eta) &= a(\eta) S_k(\chi_0) \label{otheruseful}
\\
F^2 \dot{T}^2 &= \dot{R}^2 + F a^2 ,  \label{useful}
\end{align}%
where we let $F = f(R(\eta))$. 

The normal vector field to $\Sigma^+$ (pointing into the Schwarzschild bulk) is given by
\begin{align}
n_\mu dx^\mu &= -\frac{\dot{R}}{a} dt + \frac{\dot{T}}{a} dr \label{normalcov}
\\
n^\mu \partial_\mu &= \frac{\dot{R}}{a F} \partial_t + \frac{F \dot{T}}{a} \partial_r
\end{align}
where we have used Eq.~\eqref{useful} to simplify the normalization. Given the projection tensor
\begin{align}
\tensor{e}{^\mu_i} = \frac{\partial x^\mu}{\partial y^i},
\end{align}
it is straightforward to check that $\tensor{e}{^\mu_i} n_\mu = 0$.

We now turn to the extrinsic curvature on the boundary. We wish to compute
\begin{align}
K^+_{ij} = \tensor{e}{^\mu_i} \tensor{e}{^\nu_j} \nabla_\mu n_\nu.
\end{align}
The projection tensors will project out all components perpendicular to $\Sigma^+$, but if $\nabla_{\mu}n_{\nu}$ were calculated before multiplying by $\tensor{e}{^\mu_i}\tensor{e}{^\nu_j}$, it would be necessary to extend the normal vector field off the boundary.

To avoid having to do so, we rewrite the extrinsic curvature in terms of the velocity vector $u^\mu$ for a particle traveling along a geodesic on $\Sigma^+$, 
\begin{align}
u^\mu = \frac{\partial x^\mu}{\partial \tau}= \frac{1}{a}\frac{\partial x^\mu}{\partial \eta}.
\end{align}
Hence, the components of $u^\mu$ are
\begin{align}
u^\mu \partial_\mu = \frac{1}{a} \left(\dot{T} \partial_t + \dot{R} \partial_r\right).
\end{align}
It is straightforward to check that $n_\mu u^\mu = 0$, which upon differentiation yields
\begin{align}
u^\mu \nabla_\nu n_\mu = - n_\mu \nabla_\nu u^\mu.
\end{align}

Noting that $\tensor{e}{^\mu_\eta} = a u^\mu$, we can compute
\begin{align}\label{Ketaeta}
K^+_{\eta \eta} = a^2 u^\mu u^\nu \nabla_\mu n_\nu = - a^2 u^\mu n_\nu \nabla_\mu u^\nu = - a^2 n_\nu a^\nu,
\end{align}
where we define the acceleration vector to be $a^\nu = u^\mu \nabla_\mu u^\nu$. Importantly, this is in the form of a parallel transport expression, and so we can write
\begin{align}
a^\nu = \frac{d^2x^\nu}{d\tau^2} + \Gamma^{\nu}_{\sigma \lambda} \frac{dx^\sigma}{d\tau} \frac{dx^\lambda}{d\tau}.
\end{align}
This formula could be evaluated straightforwardly, but it is easier to first rewrite it as an expression for $a_{\mu}\equiv g_{\mu\nu}a^{\nu}$, which is not very often written but which is very useful:
\begin{align}
	a_{\mu}=\frac{d}{d\tau}\left(g_{\mu\nu}\frac{dx^{\nu}}{d\tau}\right)-\frac{1}{2}\frac{\partial g_{\lambda \sigma}}{\partial x^{\mu}}\frac{dx^{\lambda}}{d\tau}\frac{dx^{\sigma}}{d\tau}.
\end{align}
In this form it is easy to see that 
\begin{align}
	a_0=-\frac{d}{d\tau}\left(F\frac{dT}{d\tau}\right),
\end{align}
since $g_{\mu\nu}$ is independent of $t$. $a_r$ is slightly more complicated, but since we know the direction of $a_\mu$, it will be sufficient for us to know a single component. From the spherical symmetry we know that $a_\mu$ has no component in the $\theta$ or $\phi$ directions, so it must lie in the $r$-$t$  plane. But it must be perpendicular to $u^{\mu}$, since $u_\mu a^\mu = u_{\mu}u^{\nu}\nabla_{\nu}u^{\mu}=\frac{1}{2}u^{\nu}\nabla_{\nu}(u_{\mu} u^{\mu})=0$. Hence it must be proportional to $n_{\mu}$, so knowledge of $a_0$, together with Eq.~\eqref{normalcov}, gives 
\begin{align}
	a_{\mu}=\left(\frac{dR}{d \tau}\right)^{-1}\frac{d}{d\tau}\left(F \frac{dT}{d\tau}\right)n_{\mu}. 
\end{align}
Since $n_{\mu}n^{\mu}=1$, from Eq.~\eqref{Ketaeta} we find 
\begin{align}
	K^{+}_{\eta\eta}=-a^2\left(\frac{dR}{d\tau}\right)^{-1}\frac{d}{d\tau}\left(F\frac{dT}{d\tau}\right).
\end{align}

The other components of $K^+_{ij}$ are simpler to compute, as due to the symmetry of the extrinsic curvature, they can always be written as derivatives of $n_\theta$ and $n_\phi$. The only nonzero terms are
\begin{align}
K^+_{\theta \theta} &= -\Gamma^{r}_{\theta\theta} n_r = a F \frac{dT}{d\tau}S_{k}(\chi_0)
\\
K^+_{\phi \phi} &= -\Gamma^{r}_{\phi\phi} n_r = a F \frac{dT}{d\tau}S_{k}(\chi_0)\sin^2\theta.
\end{align}

\subsection{Matching Conditions}\label{sub:matching}

We have already used the first Israel junction condition, which requires the induced metrics to agree. This led us to Eqs.~\eqref{otheruseful} and \eqref{useful}. 
Now that we have explicit forms for the extrinsic curvature for both sides of the hypersurface, we can apply the second Israel junction condition, which requires that
\begin{align}
K^+_{ij} = K^-_{ij}
\end{align}
in the absence of surface stress-energy. The $\eta\eta$ and $\theta\theta$ components yield two independent conditions, with the $\phi\phi$ component equivalent to the $\theta\theta$ component. Thus we have
\begin{align}
	0&= \frac{d}{d\tau}\left(F \frac{dT}{d\tau}\right),\\
	S'_{k}(\chi_0)&=F \frac{dT}{d\tau}. \label{eq:matching}
\end{align}
The first of these equations clearly follows as a consequence of the second, so we need only enforce the second equation. Squaring the equation and using Eq.~\eqref{useful}, one has 
\begin{align}
S'^2_k(\chi_0)=\left(\frac{\dot{R}}{a}\right)^2+F.
\end{align}
Remembering that $R(\eta) = a(\eta) S_k(\chi_0)$, this becomes
\begin{align}
S'^2_k(\chi_0)=\left(\frac{\dot{a}}{a}\right)^2 S_k^2(\chi_0)+F.
\end{align}
We now use the Friedmann equation \eqref{friedmann} to replace $(\dot{a}/a)^2$, and substitute $F = 1 - 2M/R$.
\begin{align}
S'^2_k(\chi_0)=\left(\frac{8\pi}{3} \rho a^2-k\right)S_{k}^2(\chi_0)+1-\frac{2M}{aS_{k}(\chi_0)}.
\end{align}
Using Eq.~\eqref{Srelations}, this reduces to 
\begin{align}
	0=\frac{8\pi}{3} \rho a^2 S_{k}^2(\chi_0)-\frac{2M}{aS_{k}(\chi_0)},
\end{align}
which immediately gives us our final result for the mass of a patch of an FRW universe:
\begin{align}\label{Mass}
M = \frac{4 \pi \rho a^3}{3} S_k^3(\chi_0) = \frac{4 \pi R^3}{3} \rho.
\end{align}
This gives a very natural interpretation for the mass in terms of the circumferential Schwarzschild radius.
%

\subsection{Implications}
\label{sub:implications}
Having stitched our two regions together, we can now look at the implications. By applying the junction conditions, we found that the mass $M$ of the FRW patch can be computed using Eq.~\eqref{Mass}. This mass is the Schwarzschild mass, and hence the ADM mass of the patch.
%

For the case that we studied of a dust-filled universe (with $P=0$), $\rho \propto 1 / a^3$, and the expression for mass \eqref{Mass} remains constant under time evolution.
%
Rewriting Eq.~\eqref{Mass} for each value of $k$,
\begin{align}
M_{k}=\begin{cases}
\label{masscase}
\frac{4}{3}\pi a^3\rho\sinh^3(\chi_0), &\text{for } k=-1\\
\frac{4}{3}\pi a^3\rho\chi_0^3,        &\text{for } k=0\\
\frac{4}{3}\pi a^3\rho\sin^3(\chi_0),  &\text{for } k=+1.
\end{cases}
\end{align}
We can compute the volume of the FRW patch by integrating
\begin{align}\label{relvol}
V_k = \int \sqrt{|h|}d^3\xi = \int a^3 S_{k}^2(\chi) \sin(\theta) d\chi d\theta d\phi
\end{align}
where $h_{ij}$ is the induced metric on the equal-time hypersurface within the patch. The three cases give
\begin{align}\label{volumes}
V_{k} = \begin{cases}
\pi a^3 \left[-2\chi_0+\sinh(2\chi_0)\right]  & \text{for } k=-1\\
\frac{4}{3} \pi a^3\chi_0^3,                  & \text{for } k=0\\
\pi a^3 \left[2 \chi_0-\sin(2\chi_0)\right] & \text{for } k=+1.
\end{cases}
\end{align}

We thus confirm the result of Zel'dovich that in the limit as $\chi_0 \rightarrow \pi$, the closed universe has zero mass, but finite volume. The open and flat cases, on the other hand, have divergent mass and volume as $\chi_0$ increases, as expected. In the flat case, the effective density is independent of $\chi_0$, $M_0/V_0 = \rho$, with mass and volume growing in equal proportions. In the open case, $M_{-1} / V_{-1} \sim \rho e^{\chi_0}$ grows exponentially with increasing $\chi_0$.



\section{Newtonian Limit}\label{sec:newtonian_cosmology}

A check on our work for small patches of FRW (small $\chi_0$) is provided by comparison to the results of classical Newtonian gravity. We should expect that the total mass of a patch given by Eq.~\eqref{Mass} in the case of dust should reduce, for small radii, to what one would calculate in Newtonian gravity.  Of course purely Newtonian physics does not allow a calculation of the rest energy, but if we define the energy of a Newtonian model of a ball of expanding gas as $M_{\mathrm{rest}} c^2$ plus the Newtonian mechanical energy (kinetic plus potential), where $M_{\mathrm{rest}}$ is the total mass, we expect agreement to the appropriate order with the fully relativistic calculation of Eq.~\eqref{Mass}.  In this section only, overdots refer to derivatives with respect to cosmological time ($dt = a d\eta$).

To compare with the model of Section \ref{sec:constructing_the_model}, we consider a Newtonian model of a uniformly expanding sphere of dust, which at some chosen time has the same volume, mass density, and instantaneous Hubble expansion rate as the relativistic model.  For a dust universe, the Newtonian model obeys exactly the same Friedmann equation as the relativistic model, although it contains no information about the spatial curvature of the relativistic model.  The volume, mass density, and Hubble expansion rate give enough information to specify an initial value problem in either the relativistic or the Newtonian model, and these quantities will evolve in exactly the same way in either model.  For the comparison, we match volumes so that the Newtonian model has the same total rest mass, or equivalently the same number of dust particles, as the relativistic model. 
Since we are interested in small $\chi_0$, we expand Eq.~\eqref{volumes} in a power series,
\begin{align}
V_k = \frac{4 \pi}{3} a^3 \chi_0^3 \left(1 - \frac{1}{5} k \chi_0^2 + \frac{2}{105} k^2 \chi_0^4 + O\left(\chi_0^6\right) \right).
\end{align}
The rest mass of the Newtonian sphere (as well as the relativistic sphere) is then
\begin{align}
M_{\mathrm{rest}} = \rho V_k .
\end{align}
Equating $V_k$ with the Newtonian volume
\begin{align}
V_k = V_{\rm N} = \frac{4 \pi}{3} R_{\rm N}^3,
\end{align}
we find that the radius of the sphere in the Newtonian model is
\begin{align}
R_{\rm N} = a \chi_0 \left(1 - \frac{1}{15} k \chi_0^2 + O\left(\chi_0^4\right) \right).
\end{align}
For the Newtonian ball of expanding dust, a dust particle at position $\vec{r} = a \chi \hat{r}$ has a velocity $\vec{v} = (\dot{a}/a) \vec{r}$ (Hubble's law). By considering a sphere of dust built up out of thin spherical shells with mass $dm = \rho dV = 4\pi \rho r^2 dr$, we find that the total kinetic energy of the sphere is given by
\begin{align}
K = \frac{2 \pi}{5} \rho R_{\rm N}^5 \left(\frac{\dot{a}}{a}\right)^2 \label{kinetic1}.
\end{align}
Transforming the Friedmann equation \eqref{friedmann} to cosmological time
\begin{align}
\left(\frac{\dot{a}}{a}\right)^2 = \frac{8\pi\rho}{3}-\frac{k}{a^2}
\end{align}
 the kinetic energy can be written as
\begin{align}
K = \frac{3}{5}\frac{M_{\mathrm{rest}}^2}{R_{\rm N}} - \frac{3}{10} k M_{\mathrm{rest}} \left(\frac{R_{\rm N}}{a}\right)^2. \label{kinetic2}
\end{align}

The mass enclosed in a sphere of radius $r$ is given by $M(r) = (4 \pi/3) r^3 \rho$, and the gravitational potential energy of a spherical shell is $dU = - M(r) dm / r$. Integrating over shells, we compute the potential energy of the sphere to be
\begin{align}
U = - \frac{3}{5} \frac{M_{\mathrm{rest}}^2 }{R_{\rm N}}. \label{potential}
\end{align}
Combining Eqs.~\eqref{kinetic2} and \eqref{potential}, we see that the potential energy is precisely canceled by the first term in \eqref{kinetic2}. Thus the mechanical energy $E_{\mathrm{mech}}=K+U$ is given 
 by
\begin{align}\label{emech}
E_{\mathrm{mech}} = - \frac{3}{10} k M_{\mathrm{rest}} \chi_0^2 \left(1 - \frac{2}{15} k \chi_0^2  + O(\chi_0^4) \right).
\end{align}
Note that this is vanishing for a flat geometry where the kinetic and potential energies exactly cancel, while the open and closed geometries, for the same value of $\chi_0$, have finite values which to lowest order are equal in magnitude but opposite in sign.

The total Newtonian energy $M^k_{\mathrm{Newt}} = M_{\mathrm{rest}} + E_{\mathrm{mech}}$ is then
\begin{align}
M^k_{\mathrm{Newt}} = M_{\mathrm{rest}} \left(1 - \frac{3}{10} k \chi_0^2 + \frac{1}{25} k^2 \chi_0^4 + O\left(\chi_0^6\right) \right).
\end{align}
For comparison, the total mass for the relativistic model, given by Eq.~(\ref{Mass}), can be expanded in the same variables, giving
\begin{align}
M^k_{\mathrm{rel}} = M_{\mathrm{rest}} \left(1 - \frac{3}{10}  k \chi_0^2 + \frac{41}{1400} k^2 \chi_0^4 + O\left(\chi_0^6\right) \right).
\end{align}
Thus, the relativistic and Newtonian energies agree on the leading order correction to the rest mass energy, as we would expect, but they disagree at the next order. To improve the matching to higher order would require including higher-order corrections in a post-Newtonian expansion.

\section{Boundary Trajectory for Dust}\label{sec:boundary_trajectory_dust}


The boundary of the FRW patch undergoes a nontrivial trajectory in Schwarzschild coordinates. In this section, we solve for the trajectory and demonstrate its evolution for open, flat and closed universes.

\begin{figure}[h]
\centering
\includegraphics[width=\columnwidth]{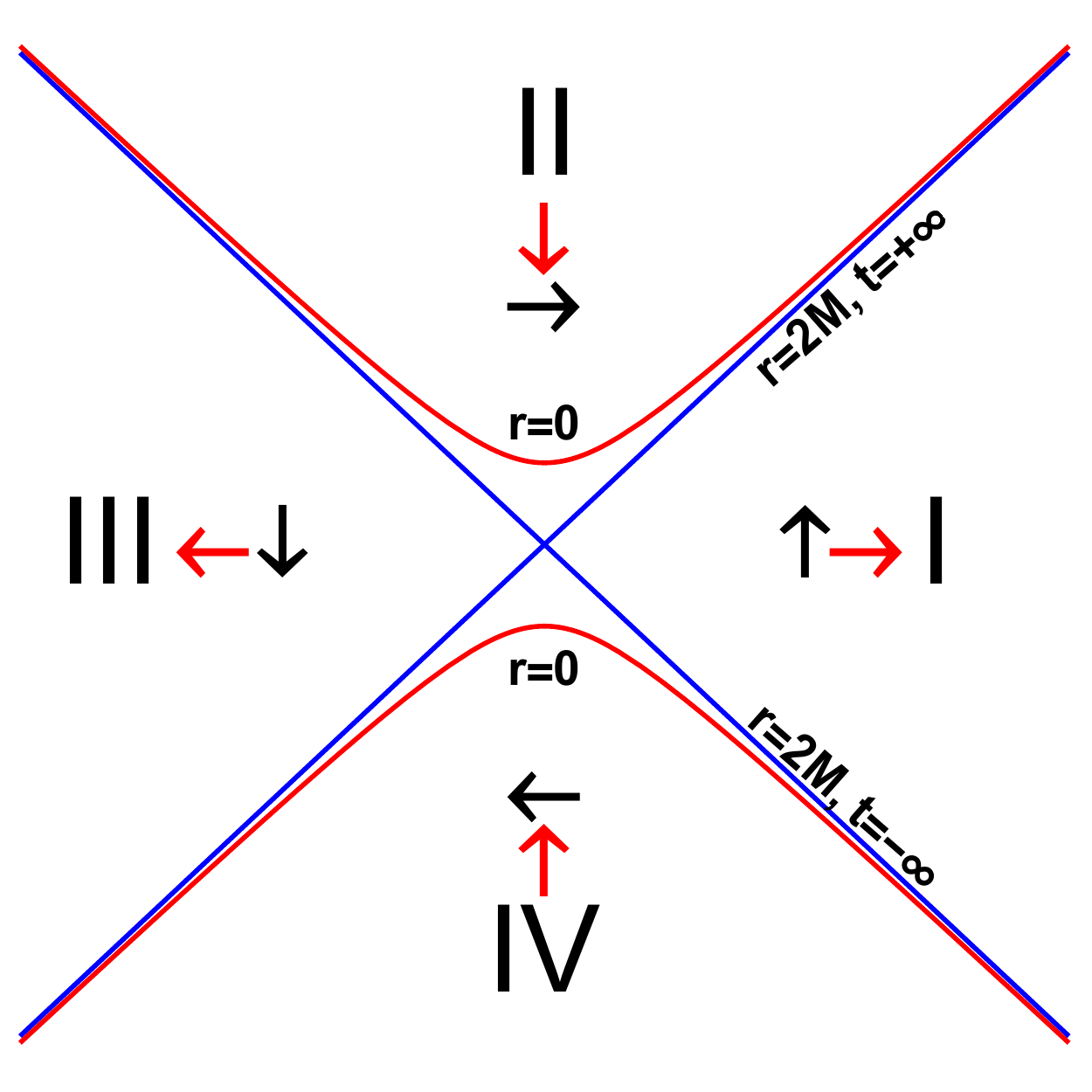}
\caption{\label{fig:kruskalconventions} This diagram shows the conventions we will use throughout the paper for labeling quadrants of a Kruskal diagram. The quadrants of a Penrose diagram are labeled in the corresponding manner. Black and red arrows indicate the directions in which the Schwarzschild time ($t$) and radial ($r$) coordinates increase, respectively.}
\end{figure}

Let us begin by understanding our coordinate systems in detail. In FRW coordinates, we have a conformal time parameter $\eta$ which is a future-directed time-like coordinate. The evolution of the boundary begins with a singularity at $a(\eta_0) = 0$, and the FRW patch subsequently grows with $\dot{a} > 0$. In Schwarzschild coordinates, the boundary can be either inside or outside the gravitational radius. In various regimes, the boundary can pass through all four regions of a Kruskal diagram (see Fig. \ref{fig:kruskalconventions}). We define the Kruskal $U$ and $V$ coordinates in the four regions by
\begin{subequations}
\begin{align}
	\label{kruskallightcoords}
	\text{Region I:} \quad U&=-\sqrt{r/2M-1}\,e^{(r-t)/4M} 
	\\ V&=\sqrt{r/2M-1}\,e^{(r+t)/4M}, \nonumber
	\\
	\text{Region II:} \quad U&=\sqrt{1-r/2M}\,e^{(r-t)/4M} 
	\\ V&=\sqrt{1-r/2M}\,e^{(r+t)/4M}, \nonumber
	\\
	\text{Region III:} \quad U&=\sqrt{r/2M-1}\,e^{(r-t)/4M} 
	\\ V&=-\sqrt{r/2M-1}\,e^{(r+t)/4M}, \nonumber
	\\
	\text{Region IV:} \quad U&=-\sqrt{1-r/2M}\,e^{(r-t)/4M} 
	\\ V&=-\sqrt{1-r/2M}\,e^{(r+t)/4M}. \nonumber
\end{align}
\end{subequations}
We can then define Kruskal ${\cal T}$ and ${\cal X}$ variables in all four regions as 
\begin{align}
	\label{kruskalcoords}
	\mathcal{T}=\frac{U+V}{2}, \qquad \mathcal{X}=\frac{V-U}{2}.
\end{align}
From these definitions, it is straightforward to show that
\begin{align}
{\cal T}^2 - {\cal X}^2 = \left(1-\frac{r}{2M}\right) e^{r/(2M)},
\end{align}
and so contours of constant $r$ are hyperbolas in the ${\cal T}$-${\cal X}$ plane, in the usual fashion. It is also straightforward to show that
\begin{align}
\tanh\left(\frac{t}{4M}\right) = \begin{cases}
\displaystyle \frac{\cal T}{\cal X} & \text{in regions I, III},
\\
\phantom{.}
\\
\displaystyle \frac{\cal X}{\cal T} & \text{in regions II, IV}
\end{cases}
\end{align}
and so lines of constant $t$ are always straight lines through the origin.

The trajectory of the boundary in Schwarzschild coordinates is parametrized by $\eta$. The radial position of the boundary varies with $\eta$ as
\begin{align}
\frac{dR}{d\eta} = \frac{\dot{a}}{a} R,
\end{align}
and so the sign of $dR/d\eta$ depends on the sign of $\dot{a}$ ($a$ and $R$ are always positive). In region IV, $R$ is time-like and future-directed, and so we must have $\dot{a} > 0$ in region IV. Similarly, in region II, we must have $\dot{a} < 0$. In regions I and III, $\dot{a}$ can take any sign.

We can solve for $dT/d\eta$ from the second Israel junction condition. Eq.~\eqref{eq:matching}. 
\begin{align}\label{dTdeta}
\frac{dT}{d\eta} = \frac{R S'_k(\chi_0)}{S_k(\chi_0)(1-2M/R)}.
\end{align}
The sign of $dT/d\eta$ depends on the sign of $S'_k(\chi_0)$ and whether $R < 2M$ or $R > 2M$. For the open, flat, and closed cases with $\chi_0 < \pi/2$, $dT/d\eta$ is negative for $R < 2M$ and positive for $R > 2M$. For the closed case with $\chi_0 = \pi/2$, $dT/d\eta = 0$, and for $\chi_0 > \pi/2$, $dT/d\eta$ is positive for $R < 2M$ and negative for $R > 2M$. Knowing these signs allows us to choose between various $\pm$ signs below.

%

We now set about solving Eq. \eqref{dTdeta}. As $a(\eta)$ is monotonically increasing/decreasing in an expansion/contraction phase, it is convenient to change variables from $\eta$ to $R(\eta)$, and write $T(R)$ piecewise for each phase. The only difference between expanding and contracting phases is an overall minus sign, which we will account for when stitching solutions together. Performing the coordinate transformation, we obtain
\begin{align}
\frac{dT}{dR} &= \pm \frac{a}{\dot{a}} \frac{R S'_k(\chi_0)}{S_k(\chi_0)(R-2M)}
\end{align}
We can write the Friedmann equation as
\begin{align}
\left(\frac{\dot{a}}{a}\right)^2 = \frac{2M}{R S^2_k(\chi_0)} - k.
\end{align}
Note this implies a maximum radius of 
\begin{align}
	\label{bounceradius}
	R_{\mathrm{max}} = 2M/\sin^2(\chi_0)
\end{align}
in the closed case. Solving for $\dot{a}/a$ and inserting into $dT/dR$ yields
\begin{align}
\frac{dT}{dR} &= \pm \frac{R}{R-2M} \sqrt{\frac{R (1 - k S^2_k(\chi_0))}{2M - k R S_k^2(\chi_0)}}
\end{align}
where we have used Eq. \eqref{Srelations}, and absorbed all sign ambiguities from square roots into the $\pm$. Finally, we define dimensionless quantities $\tilde{R} = R / M$ and $\tilde{T} = T / M$, and also write $x = S_k(\chi_0)$ for brevity.
\begin{align} \label{Tode}
\frac{d\tilde{T}}{d\tilde{R}} = \pm \frac{\tilde{R}}{\tilde{R}-2} \sqrt{\frac{\tilde{R} (1 - k x^2)}{2 - k x^2 \tilde{R}}}
\end{align}
Choosing the + sign and integrating, we obtain the following solutions.
\begin{align}
\tilde{T}_{k=0}(\tilde{R}) = \frac{\sqrt{2\tilde{R}}(6+\tilde{R})}{3} + 2 \ln \left|\frac{\sqrt{\tilde{R}} - \sqrt{2}}{\sqrt{\tilde{R}} + \sqrt{2}}\right|
\end{align}
\begin{align}
\tilde{T}_{k=-1}(\tilde{R}) &= \sqrt{1+x^2} \frac{\sqrt{\tilde{R} (2 + \tilde{R} x^2)}}{x^2}
\nonumber\\
&\quad 
+ \frac{4x^2 - 2}{x^3} \sqrt{1+x^2} \sinh^{-1}\left(\sqrt{\frac{\tilde{R} x^2}{2}}\right)
\nonumber\\
&\quad + 2 \ln \left|\frac{\sqrt{\tilde{R}(1+x^2)} - \sqrt{2+\tilde{R} x^2}}{\sqrt{\tilde{R}(1+x^2)} + \sqrt{2+\tilde{R} x^2}}\right|
\end{align}
\begin{align}
	\label{Tclosed}
\tilde{T}_{k=+1}(\tilde{R}) &= - \sqrt{1-x^2} \frac{\sqrt{\tilde{R} (2 - \tilde{R} x^2)}}{x^2}
\nonumber\\
&\quad 
+ \frac{4x^2 + 2}{x^3} \sqrt{1-x^2} \sin^{-1}\left(\sqrt{\frac{\tilde{R} x^2}{2}}\right)
\nonumber\\
&\quad + 2 \ln \left|\frac{\sqrt{\tilde{R}(1-x^2)} - \sqrt{2-\tilde{R} x^2}}{\sqrt{\tilde{R}(1-x^2)} + \sqrt{2-\tilde{R} x^2}}\right|
\end{align}
It is straightforward to confirm that these are correct through differentiation. They need to be combined in piecewise functions with appropriate signs and constants of integration to construct the full trajectory of the boundary.

All of these expressions for $\tilde{T}(\tilde{R})$ have poles at $\tilde{R}$ from the logarithmic terms, as expected to pass through the event horizon. Note that the absolute values are required to make these expressions work for both $\tilde{R} < 2$ and $\tilde{R} > 2$. Also of interest is that under $x \rightarrow ix$, we map $\tilde{T}_{k=-1} \leftrightarrow \tilde{T}_{k=+1}$. Furthermore, $\tilde{T}_{k=0}$ is the limit of $\tilde{T}_{k = \pm 1}$ as $x \to 0$.

\begin{figure}[t]
\centering
\includegraphics[width=0.8\columnwidth]{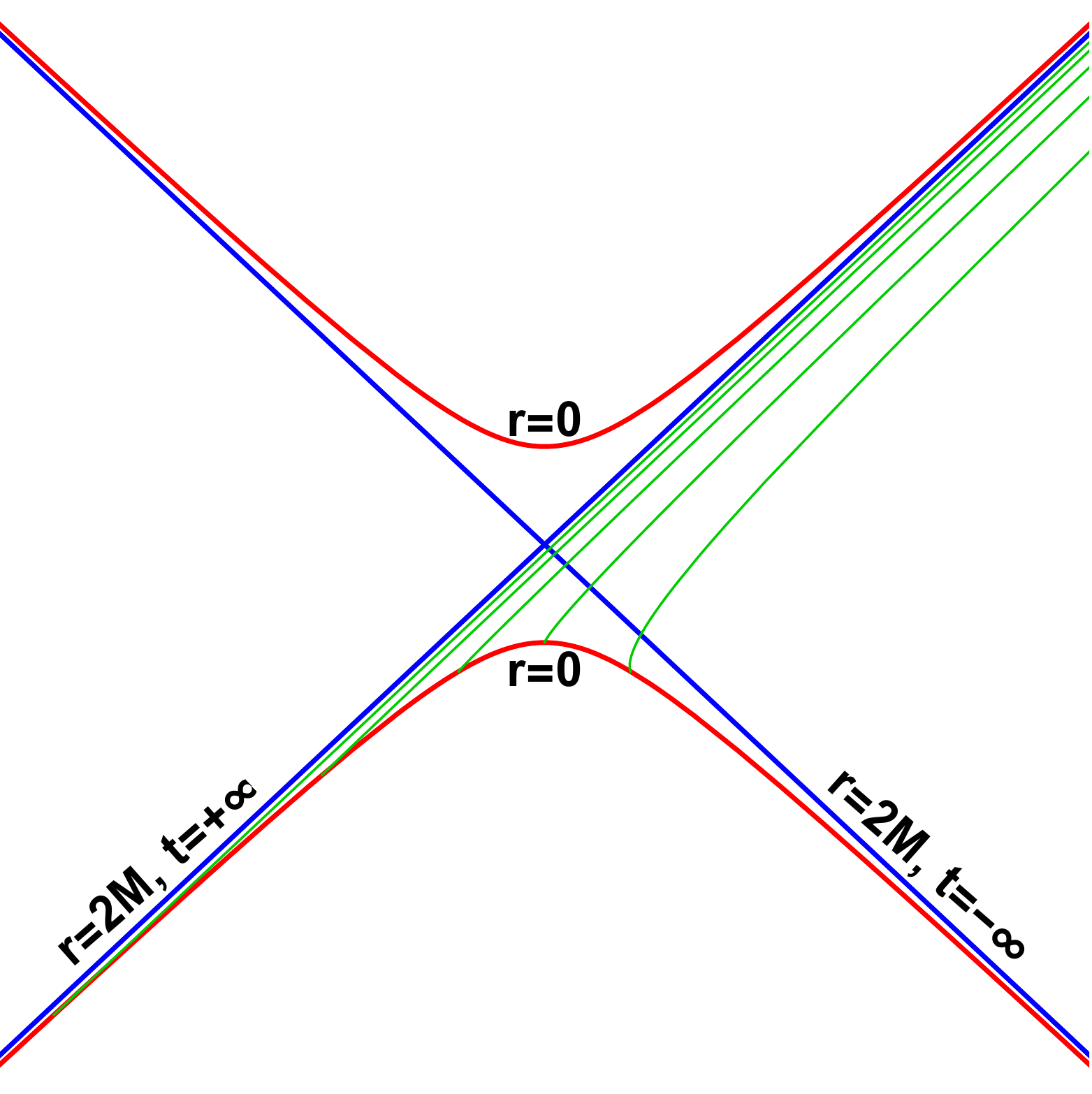}
\includegraphics[width=\columnwidth]{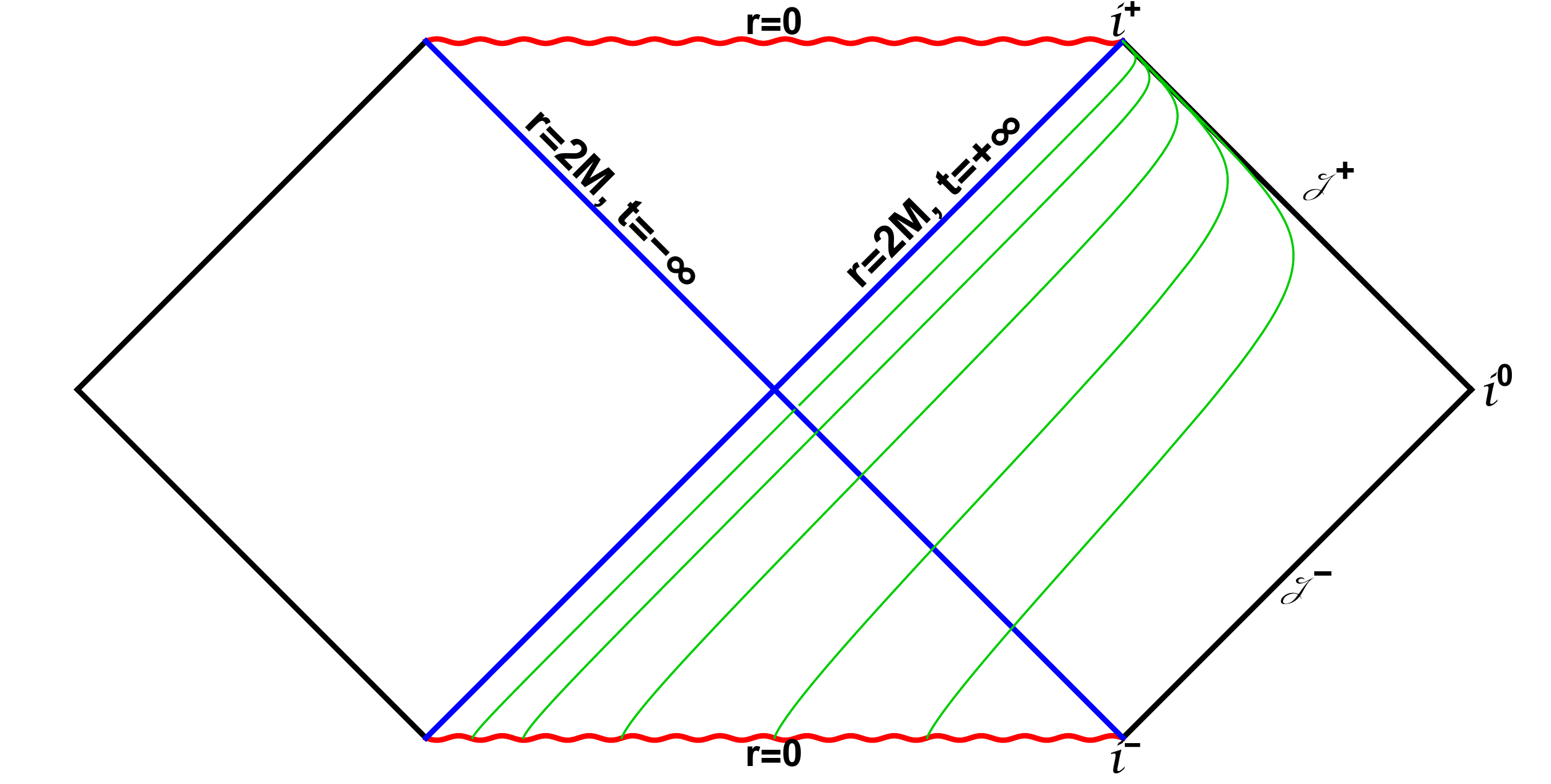}
\caption{The trajectory of the boundary of a patch for a flat FRW universe plotted in Kruskal coordinates (top) and Penrose coordinates (bottom), plotted in green. Trajectories are plotted for multiple values of the integration constant; from right to left, $\tilde{T}_0 = -3,0,3,6,9$. The trajectories start at the singularity in region IV before crossing the white hole event horizon into region I, where they expand indefinitely. Note that the trajectory converges to time-like infinity ($i^+$) rather than future null infinity ($\mathscr{I+}$).}\label{fig:flatplots}
\end{figure}

Let us start with the flat case. Here, the boundary begins in region IV, expands through the white hole event horizon, and continues expanding forever. The trajectory is given by
\begin{align}
\tilde{T}(\tilde{R}) = \tilde{T}_{k=0} (\tilde{R}) + \tilde{T}_0
\end{align}
for both $\tilde{R} < 2$ and $\tilde{R} > 2$, with constant of integration $\tilde{T}_0$. To see that the constant of integration remains the same in both regions, observe that the ratios ${\cal T}^{I} / {\cal T}^{IV}$ and ${\cal X}^{I} / {\cal X}^{IV}$ on either size of the horizon all limit to unity on the horizon. This evolution is shown on Kruskal and Penrose\footnote{We compute the Penrose coordinates $\tilde{U} = \tan^{-1}(U)$ and $\tilde{V} = \tan^{-1}(V)$, and then construct $\tilde{\cal T}$ and $\tilde{\cal X}$ from these in the same manner as in Eq. \eqref{kruskalcoords}.} diagrams in Fig.\ref{fig:flatplots}. Note that the region to the left of each trajectory is not described by the Schwarzschild metric, as it is inside the FRW patch (not represented). 

\begin{figure}
\centering
\includegraphics[width=0.8\columnwidth]{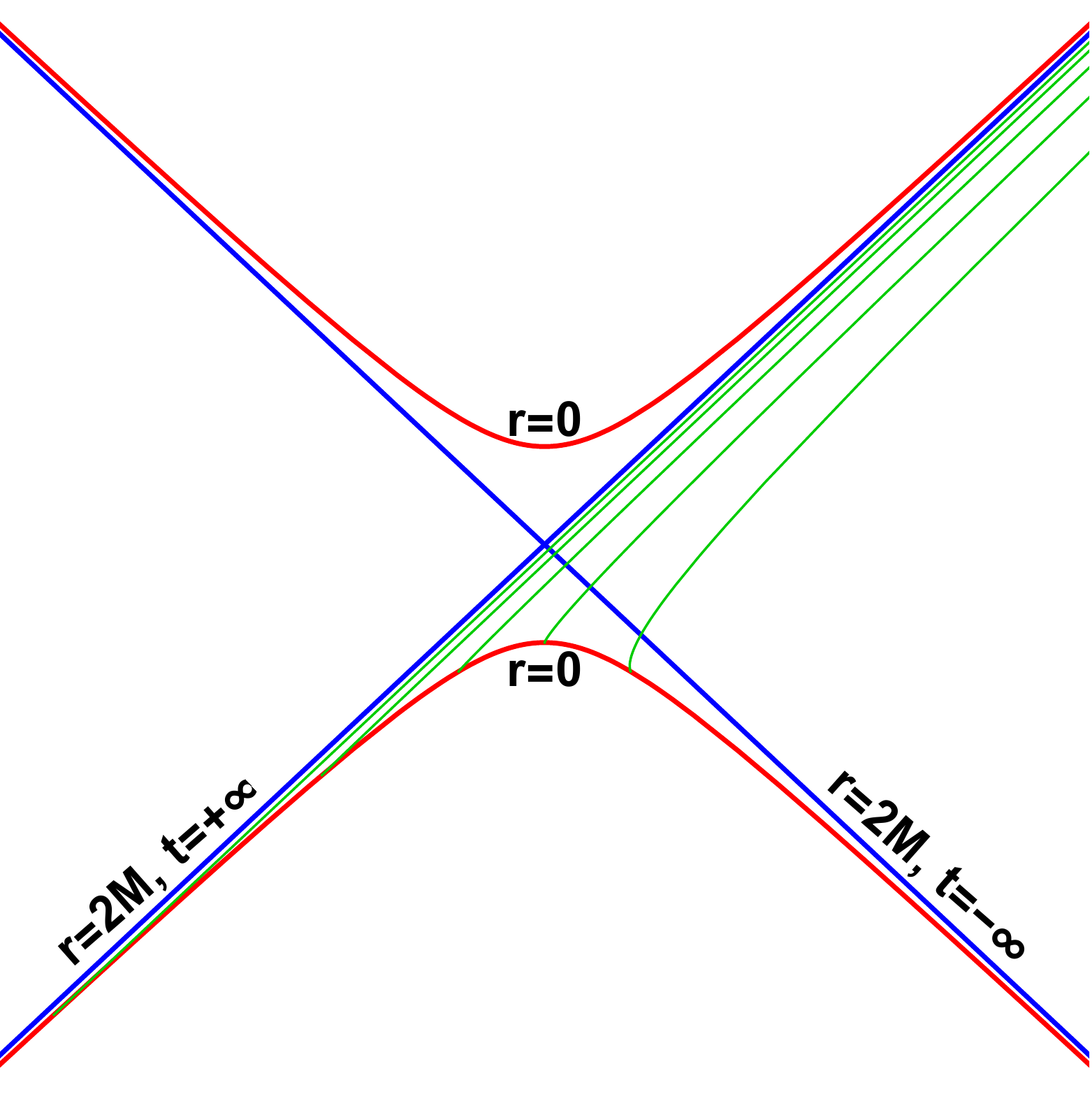}
\includegraphics[width=\columnwidth]{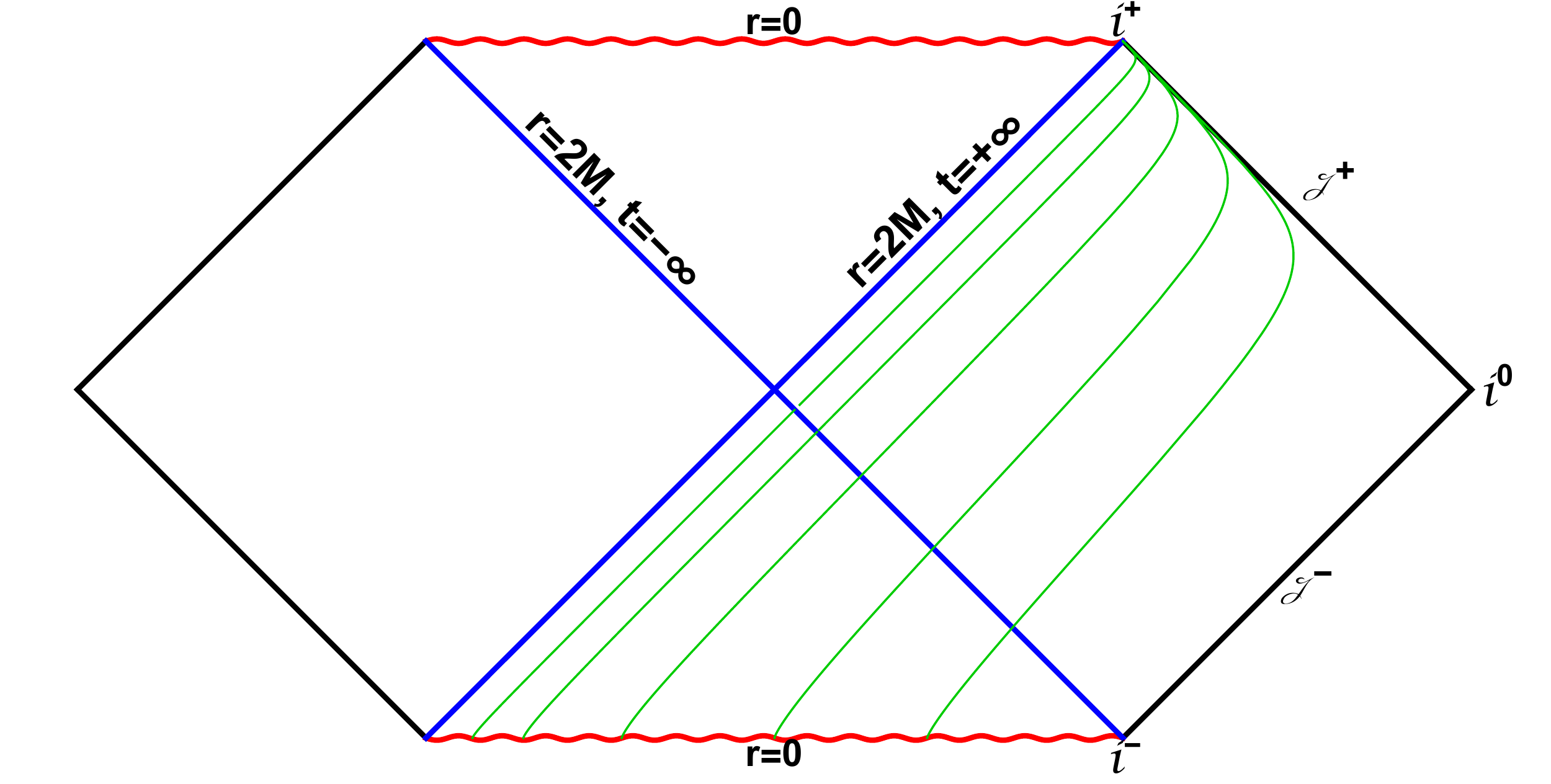}
\caption{A Kruskal diagram (top) and Penrose diagram (bottom) for a patch of an open FRW universe with $x=0.1$. Trajectories are plotted for multiple values of $\tilde{T}_0$; from right to left, $\tilde{T}_0= -3,0,3,6,9$.}\label{fig:openplots}
\end{figure}

\begin{figure}
\centering
\includegraphics[width=\columnwidth]{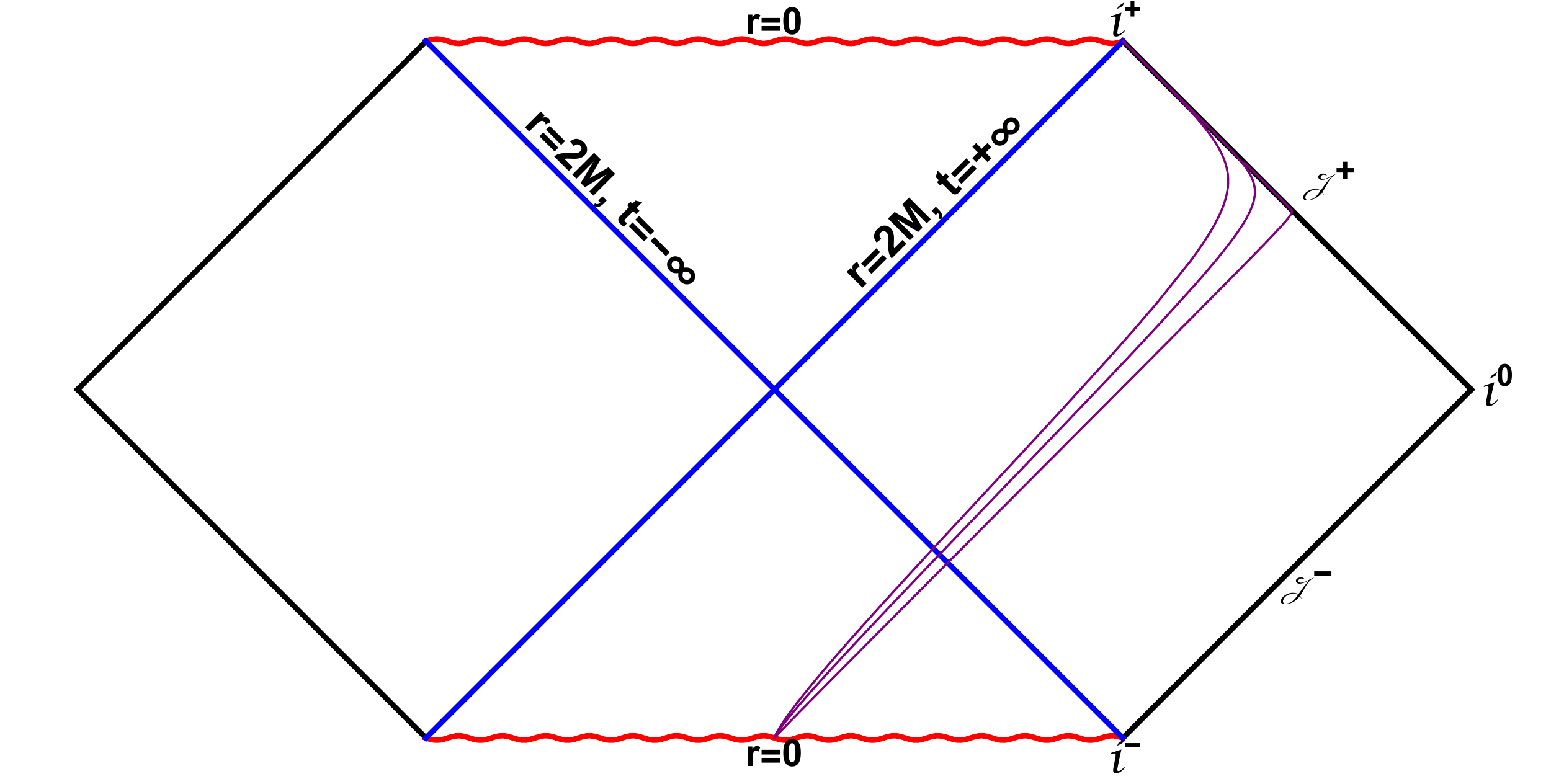}
\caption{Plotted are boundary trajectories for the open FRW patch  on a Penrose diagram with, from right to left, $x=5, 1, .001$.}\label{fig:xvalsopenpenrose}
\end{figure}

The open case is very similar to the flat case. The FRW patch begins in region IV, crosses the event horizon into region I, and continues to expand indefinitely. The position of the boundary is given by
\begin{align}
\tilde{T}(\tilde{R})=\tilde{T}_{k=-1}(\tilde{R})+\tilde{T}_0
\end{align}
with the same constant of integration in both regions. The Kruskal and corresponding Penrose diagrams for varying $\tilde{T}_0$ are shown in Fig. \ref{fig:openplots}, while Fig. \ref{fig:xvalsopenpenrose} shows the Penrose diagram for varying $x$.

\begin{figure}
\centering
\includegraphics[width=0.8\columnwidth]{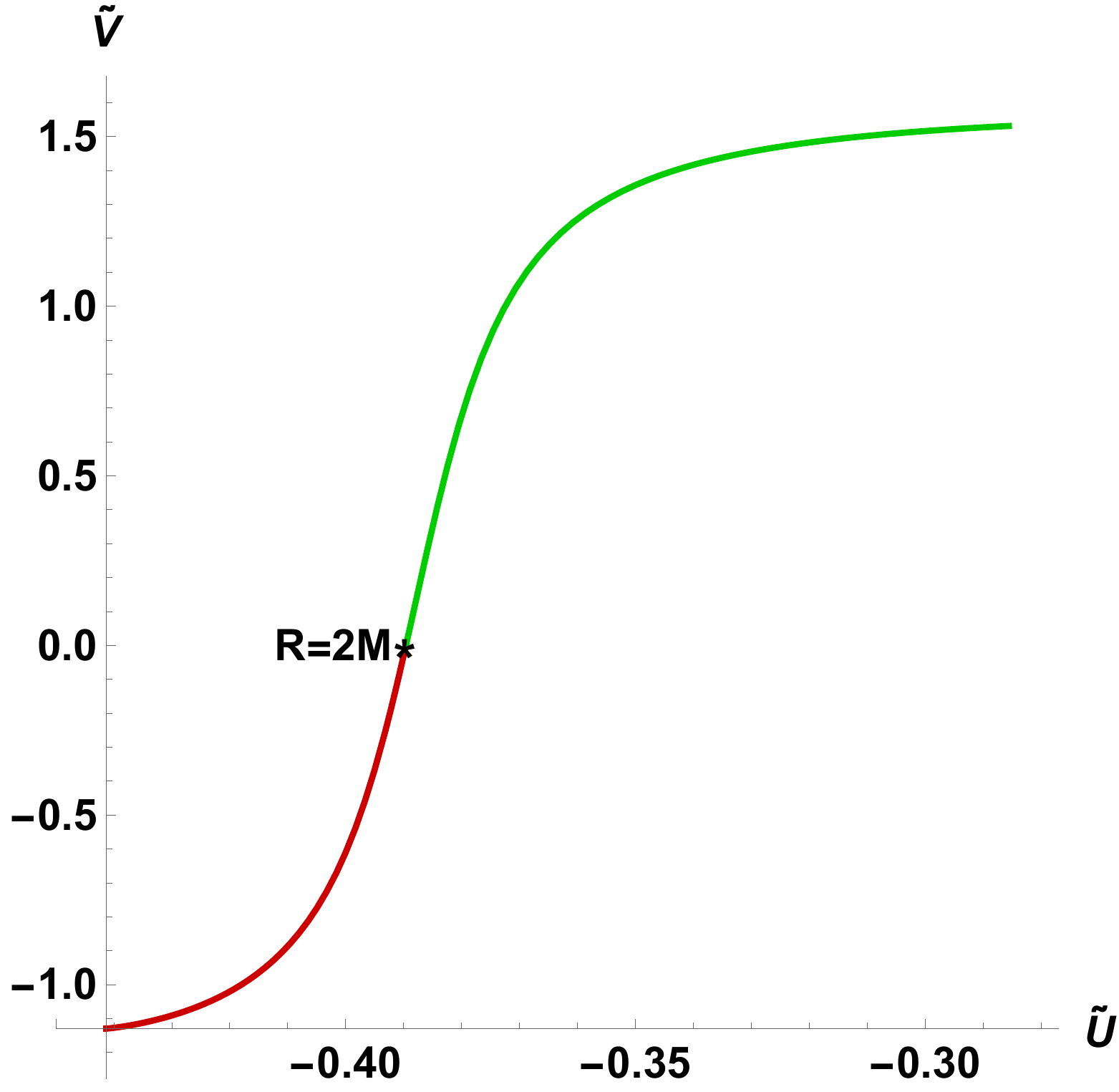}
\caption{A plot of Penrose coordinates $\tilde{V}(\tilde{U})$ for the boundary in the region $0 < R < 2M$ (red) and $2M < R < 4M$ (green) for the flat case, with $\tilde{T}_0 = 3$. Note the different scales; $\tilde{V}$ changes very rapidly through almost its entire range, while $\tilde{U}$ changes only very slightly. This is still subluminal; luminal propagation speeds would require a vertical tangent.}\label{fig:UV}
\end{figure}

In the flat and open cases, both $R$ and $T$ grow without bound. In particular, this means that the trajectory must cross every line of constant radius in the Penrose diagram. Hence, the only way to approach timelike infinity is along $\mathscr{I}^+$. We also see that the trajectory crosses the white hole event horizon at almost 45 degrees. It turns out that $\tilde{T}(\tilde{R}) \sim \sqrt{2} \tilde{R}$ at large $\tilde{R}$, and so $V$ grows exponentially with $\tilde{R}$, while $U$ is much more subdued. Hence, when flattened by the conformal map, the trajectory appears to cross the event horizon very close to 45 degrees. A plot of $\tilde{V}(\tilde{U})$ shows that while the expansion is rapid, it is still subluminal; see Fig. \ref{fig:UV}

We can investigate the speed at which the boundary expands in Schwarzschild space. The coordinate velocity is given by
\begin{align}
\frac{dR}{dT} = f(R) \sqrt{\frac{2M - k x^2 R}{R (1 - k x^2)}}
\end{align}
for $R > 2M$, which we can compare to the coordinate velocity of light,
\begin{align}
v_{null}(r) = f(r).
\end{align}
Note that as we approach $r = 2M$, both coordinate velocities are zero, as the Schwarzschild coordinate time is $-\infty$, and very small changes in radius take a long time. We stress that this is a coordinate artifact. Taking the ratio of the two velocities, we see that
\begin{align}
\frac{1}{v_{null}} \frac{dR}{dT} = \sqrt{\frac{2M - k x^2 R}{R (1 - k x^2)}}.
\end{align}
In the flat case, this decays to zero as $R \rightarrow \infty$, while in the open case, it asymptotes to $\sqrt{x^2/(1+x^2)}$.

We now turn to the boundary trajectory of a closed FRW patch. As previously noted, the closed patch will expand to a finite maximum radius, $R_{\mathrm{max}}$. It will then begin to contract, before eventually terminating at the singularity in region II. However, which coordinate patches it passes through depend on the value of $\chi_0$ chosen to define the boundary. For $\chi_0<\pi/2$, the FRW patch begins in region IV, enters region I and expands until it reaches its maximum radius. It then begins to contract and enters region II where it eventually hits the singularity. For $\chi_0>\pi/2$, the FRW patch again begins in region IV, but never emerges into the exterior Schwarzschild spacetime. Instead it expands, entering region III, then begins to contract and crosses into region II where it reaches the singularity. This is the case that Zel'dovich called the ``semi-closed world'', and possesses a truly fascinating embedding diagram \cite{zeldovichppr}. We will address later the special case when $\chi_0$ is exactly $\pi/2$.

\begin{figure}
\centering
\includegraphics[width=0.8\columnwidth]{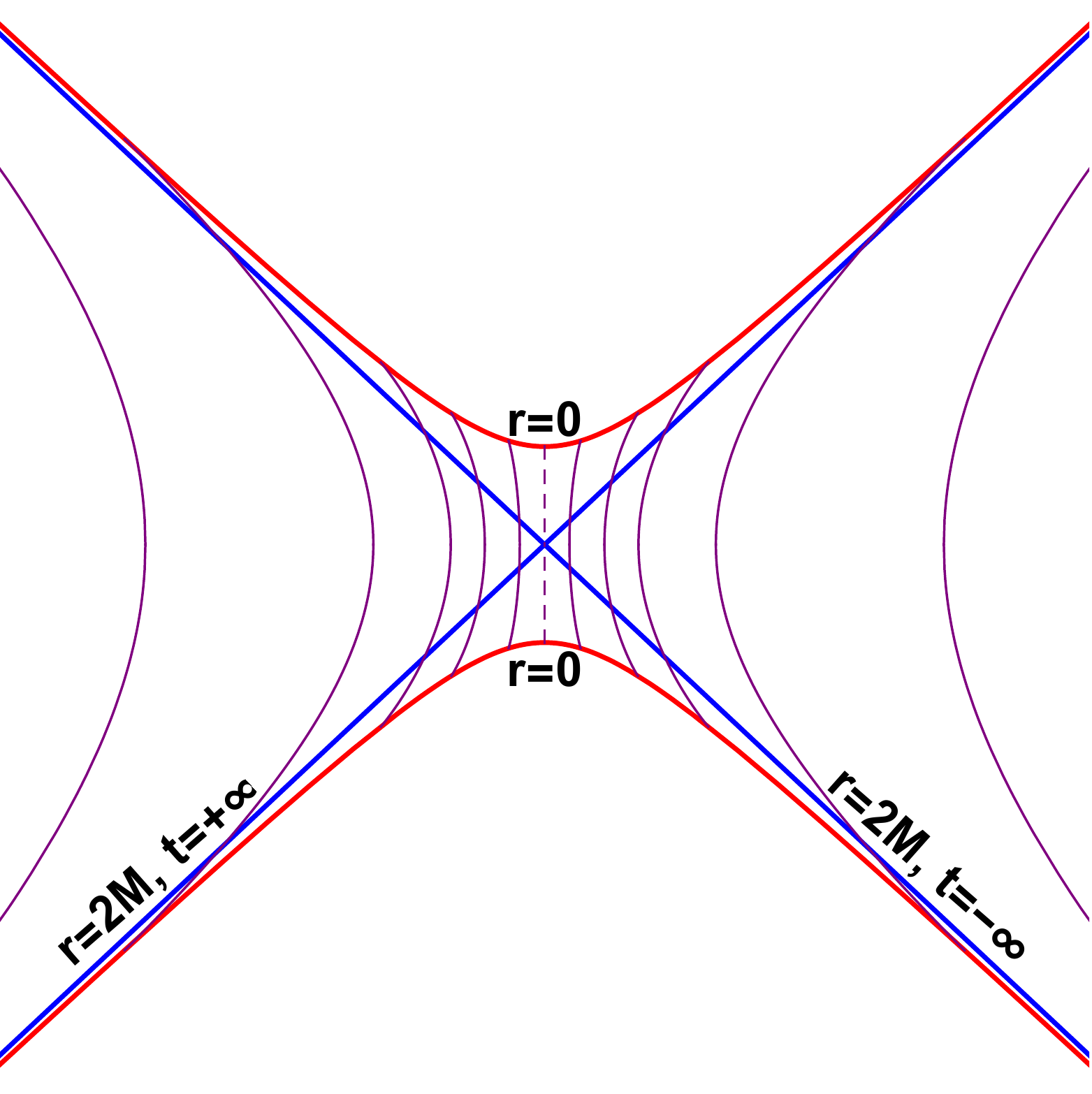}
\includegraphics[width=\columnwidth]{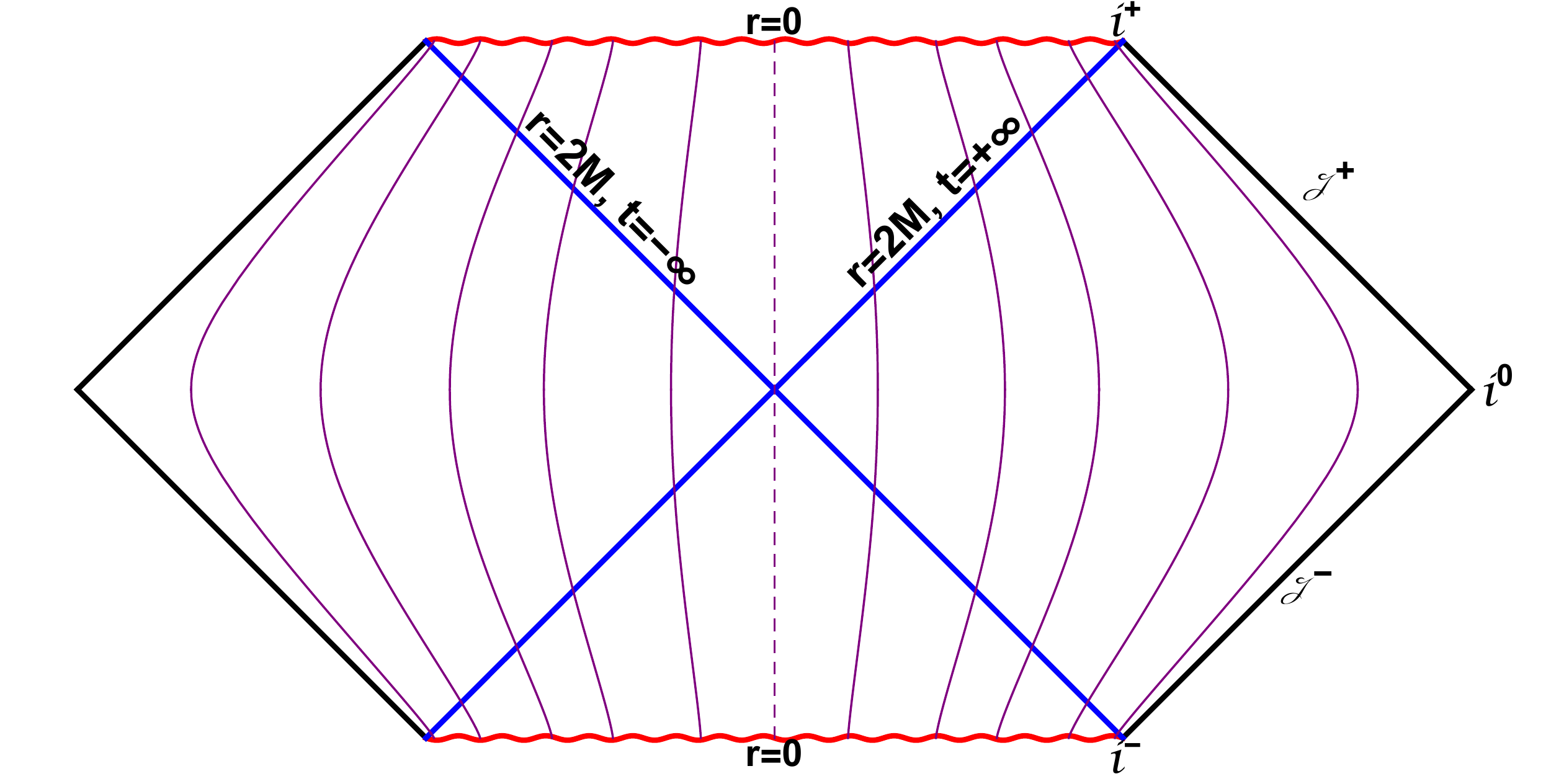}
\caption{Trajectories of the boundary of a closed FRW patch, plotted on a Kruskal diagram (top) and Penrose diagram (bottom). The special case of $x=1$ is shown for $\tilde{T}_0=0$ as the dashed curve extending directly from Region IV to Region II. For $\chi_0<\pi/2$ and $\chi_0>\pi/2$, trajectories are plotted from center to right and center to left, respectively, for $x=.99,.95,.90,.80,.65$. Note the time-reversal symmetry, which reflects the symmetry of FRW expansion and collapse in a closed universe.}\label{fig:closedplots}
\end{figure}


Before describing the trajectory, it is convenient to gauge-fix $\tilde{T} = 0$ when $R = R_{max}$, which highlights the symmetry between the expanding and contracting phases, and also simplifies stitching the two phases together. Define
\begin{align}
\tilde{T}_{sym}(\tilde{R}) = \tilde{T}_{k = 1}(\tilde{R}) - \frac{\pi\sqrt{1-x^2}(1+2x^2)}{x^3}
\end{align}
which vanishes for $\tilde{R} = \tilde{R}_{max} = 2/x^2$.

The full expression for $\tilde{T}(\tilde{R})$ must be given piecewise to join the expanding and contracting phases. For the expanding phase,
\begin{subequations}
	\begin{align}
		&\text{For} \,\,\chi_0<\frac{\pi}{2}:\ \tilde{T}(\tilde{R})=\tilde{T}_{\mathrm{sym}}(\tilde{R}),\label{smallexpand}\\
		&\text{For} \,\,\chi_0>\frac{\pi}{2}:\ \tilde{T}(\tilde{R})=-\tilde{T}_{\mathrm{sym}}(\tilde{R}),\label{largeexpand}
	\end{align}
\end{subequations}
and for the contracting phase, 
\begin{subequations}
	\begin{align}
		&\text{For} \,\,\chi_0<\frac{\pi}{2}:\ \tilde{T}(\tilde{R})=-\tilde{T}_{\mathrm{sym}}(\tilde{R}),\label{smallcontract}\\
		&\text{For} \,\,\chi_0>\frac{\pi}{2}:\ \tilde{T}(\tilde{R})=\tilde{T}_{\mathrm{sym}}(\tilde{R}).\label{largecontract}
	\end{align}
\end{subequations}
The relative sign between Eqs. \eqref{smallexpand}, \eqref{largeexpand} and Eqs. \eqref{smallcontract}, \eqref{largecontract}, is accounted for by the flip in sign of $S'(\chi_0)$ in Eq. \eqref{dTdeta} as $\chi_0$ crosses $\pi/2$.

We now return to the special case of $\chi_0=\pi/2$. In this case, $x = 1$, and from Eq. \eqref{Tode}, $\tilde{T} = \tilde{T}_0$. Such a patch has a maximum radius of $\tilde{R}_{max} = 2$, and so will never emerge into either regions I or III but will cross directly from region II into region IV. Gauge-fixing in the same way as previously chooses $\tilde{T}_0 = 0$. Kruskal and Penrose diagrams for the closed universe boundary evolution with gauge-fixed time coordinate but varying values of $\chi_0$ are shown in Fig. \ref{fig:closedplots}.

\section{Conclusions}\label{sec:conclusions}

In this paper we have extended the model put forth by Zel'dovich to an FRW universe of arbitrary curvature. We have confirmed his calculation that the total mass of a closed universe is vanishing and have found a general expression for the mass of a patch of an FRW universe. This result implies that the mass of a patch of an open or flat FRW universe grows without bound as the circumferential radius of the patch is increased. Our results indicate that even a very small patch of closed space with a large radius of curvature ($x\ll 1$) will eventually collapse to a black hole in the far distant future. One can think of the limit of $k=0$ as representing the point at which the contracting phase will take infinite time to begin. 

Back in 1963, Zel'dovich posed the challenge to extend his results for the semi-closed universe away from spherical symmetry and questioned whether closedness was a sufficient condition for precipitating the collapse of an FRW universe to a black hole. It is curious that even after fifty years, we still do not know the answers to many of the questions he posed in his conclusions. To Zel'dovich's queries we add the question of how to further generalize the model we have considered to spacetimes dominated not by dust, but by fluids with non-zero pressure, as our own universe is believed to be. Further generalization of this model might extend the
calculation of total mass of a curved spacetime to spacetimes filled with any perfect fluid. We anticipate that doing so will require a different formalism than the Israel junction conditions. Solving the Einstein constraint equations directly under spherical symmetry on a single coordinate chart should yield an expression for the mass of a patch, but will say nothing about the time evolution of the system.


\acknowledgments

This work is supported in part by the U.S. Department of Energy under grant Contract Number DE-SC0012567.

\bibliographystyle{apsrev}
\bibliography{main}

\appendix

\section{Solutions to the Friedmann equation for \texorpdfstring{$w =$}{w =} constant}\label{app:friedmann}

In this appendix, we present solutions to the Friedmann equation for an equation of state $P = w \rho$ with constant $w$. For the metric \eqref{eq:FRW}, the Friedmann equation is given by
\begin{align}
\left(\frac{\dot{a}}{a}\right)^2 = \frac{8 \pi}{3} \rho a^2 - k
\end{align}
where dots indicate derivatives with respect to conformal time, $\eta$. Note that $a$ is dimensionful. The continuity equation is given by
\begin{align}
\dot{\rho} = - 3 \frac{\dot{a}}{a} (\rho + P).
\end{align}
Assuming a constant value of $w = P / \rho$, the continuity equation can be solved to obtain
\begin{align}
\rho(a) = \rho_0 \left(\frac{a_0}{a}\right)^{3(1+w)}
\end{align}
where an arbitrary constant $a_0$ is necessary to account for the dimensions of $a$. Given this, the Friedmann equation can be written as
\begin{align}
\left(\frac{\dot{a}}{a}\right)^2 = \frac{A}{a^{1+3w}} - k
\end{align}
where
\begin{align}
A = \frac{8 \pi \rho_0 a_0^{3(1+w)}}{3}.
\end{align}
This equation is separable, and we can compute
\begin{align}
\eta - \eta_0 &= \int \frac{da}{\sqrt{A a^{1-3w} - k a^2}}
\end{align}
where we assume the universe is expanding (positive root). For open universes ($k = -1$), this gives
\begin{align}
\eta - \eta_0 &= \frac{2}{1 + 3w} \sinh^{-1} \left(\frac{1}{\sqrt{A}} a^{(1+3w)/2}\right)
\\
a(\eta) &= \left[\sqrt{A} \sinh\left(\frac{1+3w}{2} (\eta - \eta_0)\right)\right]^{2/(1+3w)}.
\end{align}
For flat universes ($k = 0$), we have
\begin{align}
\eta - \eta_0 &= \frac{2}{\sqrt{A}(1 + 3w)} a^{(1+3w)/2}
\\
a(\eta) &= \left(\frac{\sqrt{A}(1+3w)}{2} (\eta - \eta_0)\right)^{2/(1+3w)}.
\end{align}
For closed universes ($k = +1$), the solution is
\begin{align}
\eta - \eta_0 &= \frac{2}{1 + 3w} \sin^{-1} \left(\frac{1}{\sqrt{A}} a^{(1+3w)/2}\right)
\\
a(\eta) &= \left[\sqrt{A} \sin\left(\frac{1+3w}{2} (\eta - \eta_0)\right)\right]^{2/(1+3w)}.
\end{align}

\section{Alternative Derivation of the Extrinsic Curvature}\label{app:extrinsic}

In this appendix, we present an independent computation of the extrinsic curvature of the FRW boundary using Gaussian normal coordinates.

Let our coordinates in FRW be written as $\xi^\mu = (\eta, \chi, \theta, \phi)$. The boundary $\Sigma$ is described by $\chi = \chi_0$, and has coordinates $y^i = (\eta, \theta, \phi)$. We wish to transform to a Gaussian normal coordinate system given by $\xi^{\mu'} = (\bar{\eta}, z, \theta, \phi)$. We construct this coordinate system by first demanding that a point on the boundary $\xi^{\mu'} = (\bar{\eta}, 0, \theta, \phi)$ corresponds to $y^i = (\bar{\eta}, \theta, \phi)$. The coordinate $z$ describes points that move off the boundary, starting from some $y^i$, along a spatial geodesic with initial tangent vector normal to the boundary, such that $z$ is the proper distance along that geodesic.

As $\theta$ and $\phi$ are unchanged in moving to Gaussian normal coordinates, this leaves us with the task of determining $\eta(\bar{\eta}, z)$ and $\chi(\bar{\eta}, z)$ to transform the metric \eqref{eq:FRW} from FRW coordinates. In order to ensure that geodesics moving in the $\partial_z$ direction are perpendicular to the boundary, we require
\begin{align} \label{cond1}
g_{z\bar{\eta}} = a^2 \left( \frac{\partial \chi}{\partial \bar{\eta}} \frac{\partial \chi}{\partial z} - \frac{\partial \eta}{\partial \bar{\eta}} \frac{\partial \eta}{\partial z}\right) = 0.
\end{align}
To ensure that $z$ measures the proper distance along the geodesic, we require
\begin{align} \label{cond2}
g_{zz} = a^2 \left( \left(\frac{\partial \chi}{\partial z}\right)^2 - \left(\frac{\partial \eta}{\partial z}\right)^2 \right) = 1.
\end{align}

We will compute $\eta$ and $\chi$ to second order in $z$, which will be sufficient for computing the extrinsic curvature.
\begin{subequations}
\begin{align}
\eta &= \bar{\eta} + \frac{\partial \eta}{\partial z}\bigg|_{z=0} z + \frac{1}{2} \frac{\partial^2 \eta}{\partial z^2}\Bigg|_{z=0} z^2 + O(z^3)
\\
\chi &= \chi_0 + \frac{\partial \chi}{\partial z}\bigg|_{z=0} z + \frac{1}{2} \frac{\partial^2 \chi}{\partial z^2}\bigg|_{z=0} z^2 + O(z^3)
\end{align}
\end{subequations}
Using these expansions, we see that
\begin{align}
\frac{\partial \chi}{\partial \bar{\eta}} \bigg|_{z=0} = 0 \quad \text{and} \quad 
\frac{\partial \eta}{\partial \bar{\eta}} \bigg|_{z=0} = 1.
\end{align}
Using these in Eq. \eqref{cond1} evaluated at $z=0$ then requires
\begin{align}
\frac{\partial \eta}{\partial z} \bigg|_{z=0} = 0.
\end{align}
We can then evaluate Eq. \eqref{cond2} at $z=0$ to obtain
\begin{align}
\frac{\partial \chi}{\partial z} \bigg|_{z=0} = - \frac{1}{a(\bar{\eta})}
\end{align}
where we choose the negative root, as $\chi$ should decrease with increasing $z$.

To obtain the second derivatives of $\eta$ and $\chi$, we turn to the geodesic equation. Consider a curve parametrized by $z$ as $\xi^\mu(z)$. The geodesic equation that this curve satisfies can be written as
\begin{align}
\frac{d}{dz} \left(g_{\mu \rho} \frac{d\xi^\rho}{dz}\right) = - \frac{1}{2} \frac{dg_{\nu \sigma}}{d\xi^\mu} \frac{d\xi^\nu}{dz} \frac{d\xi^\sigma}{dz}.
\end{align}
The equation for $\mu = 0$ yields
\begin{align}
\frac{d^2\eta}{dz^2}
= 
\frac{1}{a(\eta)} \frac{da(\eta)}{d\eta} \left[\left(\frac{d\chi}{dz}\right)^2 - 3 \left(\frac{d\eta}{dz}\right)^2\right].
\end{align}
Evaluating this at $z=0$ gives
\begin{align}
\frac{d^2\eta}{dz^2} \bigg|_{z=0}
= 
\frac{1}{a^3(\bar{\eta})} \frac{da(\eta)}{d\eta} \bigg|_{\eta = \bar{\eta}}. \end{align}
The $\mu = 1$ equation gives
\begin{align}
\frac{d}{dz} \left(a^2(\eta) \frac{d\chi}{dz}\right) = 0,
\end{align}
which evaluated on the boundary yields
\begin{align}
\frac{d^2\chi}{dz^2} \bigg|_{z=0} = 0.
\end{align}
Hence, the coordinate transformation to Gaussian normal coordinates is given by
\begin{subequations}
\begin{align}
\eta &= \bar{\eta} + \frac{1}{2 a^3(\bar{\eta})} \frac{da}{d\eta}\bigg|_{\eta = \bar{\eta}} z^2 + O(z^3)
\\
\chi &= \chi_0 - \frac{z}{a(\bar{\eta})} + O(z^3).
\end{align}
\end{subequations}
The metric in Gaussian normal coordinates is then
\begin{align}
ds^2 &= dz^2 + a^2(\bar{\eta}) \left[- d\bar{\eta}^2 + S_k^2\left(\chi_0 - \frac{z}{a(\bar{\eta})}\right) d\Omega^2\right]
\nonumber\\
&\qquad + O(z^2).
\end{align}
On $\Sigma^-$, this reduces to the induced metric given by Eq. \eqref{induced}.

The extrinsic curvature tensor on the boundary is particularly simple in Gaussian normal coordinates.
\begin{align}
K_{ij} = - \frac{1}{2} \frac{\partial}{\partial z} g^{GN}_{ij} \bigg|_{z=0}
\end{align}
The minus sign accounts for having the normal point outwards. Correspondingly, we find
\begin{subequations}
\begin{align}
K_{\eta \eta} &= 0
\\
K_{\theta \theta} &= a(\eta) S_k(\chi_0) S'_k(\chi_0)
\\
K_{\phi \phi} &= \sin^2(\theta) K_{\theta\theta}.
\end{align}
\end{subequations}
These agree with the results in Eq. \eqref{frwextrinsic}.

\end{document}